\documentclass[10pt,emptycopyrightspace]{ewsn-proc}

\usepackage{graphicx}
\usepackage{balance}
\usepackage{comment}
\usepackage{algorithmic}
\usepackage{algorithm}
\usepackage{amsmath}
\usepackage{subfig}
\usepackage{fancyhdr}

\DeclareMathOperator*{\argmax}{arg\,max}
\DeclareMathOperator*{\argmin}{arg\,min}
\numberofauthors{1}
\author{
\alignauthor{Khaled Q. Abdelfadeel, Victor Cionca, Dirk Pesch}\\
\affaddr{Nimbus Centre, Cork Institute of Technology, Ireland}\\
\email{khaled.abdelfadeel@mycit.ie, victor.cionca@cit.ie, dirk.pesch@cit.ie}
}
\title{A Fair Adaptive Data Rate Algorithm for LoRaWAN}

\begin{document}
\maketitle
\begin{abstract}
LoRaWAN exhibits several characteristics that can lead to an unfair distribution of the Data Extracted Rate (DER) among nodes. Firstly, the capture effect leads to a strong signal suppressing a weaker signal at the gateway and secondly, the spreading codes used are not perfectly orthogonal, causing packet loss if an interfering signal is strong enough. In these conditions, nodes experiencing higher attenuation are less likely to see their packets received correctly. We develop FADR, a Fair Adaptive Data Rate algorithm for LoRaWAN that exploits the different Spreading Factors (SFs) and Transmission Powers (TPs) settings available in LoRa to achieve a fair Data Extraction Rate among all nodes while at the same time avoiding excessively high TPs. Simulations show that FADR, in highly congested cells, achieves 300\% higher fairness than the minimum airtime allocation approach and 22\% higher fairness than Brecht’s approach, while consuming almost 22\% lower energy.
\end{abstract}

%
%

%

\section{Introduction}
\label{sec:intro}

Depending on the specific radio communication conditions, a LoRaWAN gateway can decode one, all, or none of the colliding packets transmitted by multiple nodes.
When two signals using the same spreading factor (SF) arrive at the same time, with one signal stronger than the other by a certain threshold, the \textit{capture effect} causes the stronger signal to drown the weaker. This was verified experimentally by~\cite{bor2016lora}. Even when the signals use different spreading factors, this effect can still be observed, because the spreading codes are not perfectly orthogonal. However, if the power difference is below the Co-channel Interference Rejection (CIR) threshold, both signals will be decoded, whereas with the capture effect, none will. Croce \textit{et. al.} measured the CIR for LoRa in~\cite{croce2017impact}.
Furthermore, the SF allocation is affecting the probability of collision as each SF has a different airtime, e.g. using a lower SF leads to a shorter airtime which results in a lower probability of collision.

\begin{figure}
\vspace{-2.1em}
  \centering
  \subfloat[Fairness Index]{
    \label{fig:fairness_study}
    \includegraphics[width=0.25\textwidth]{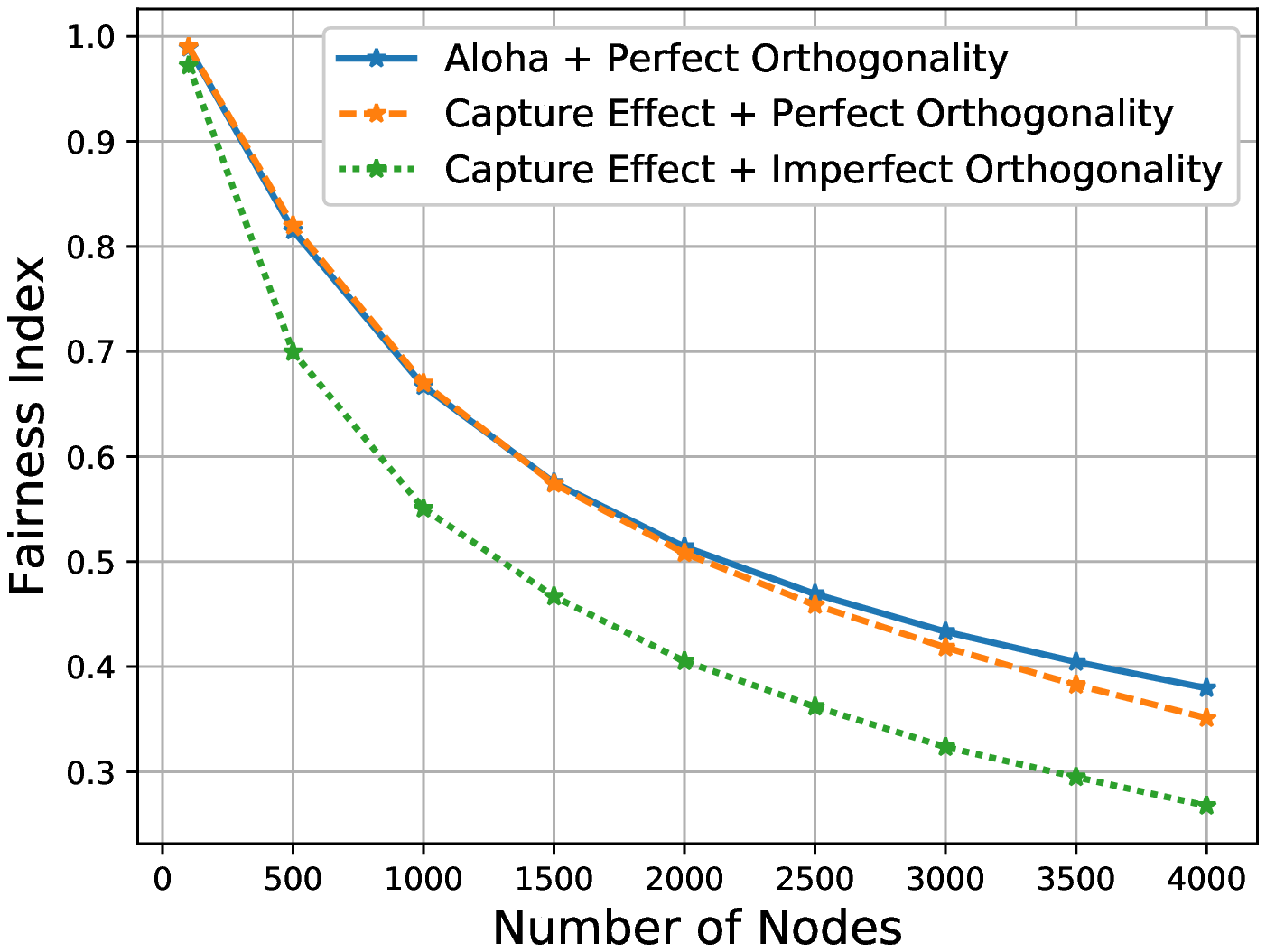}
  }
  \subfloat[DER]{
    \label{fig:der_study}
    \includegraphics[width=0.25\textwidth]{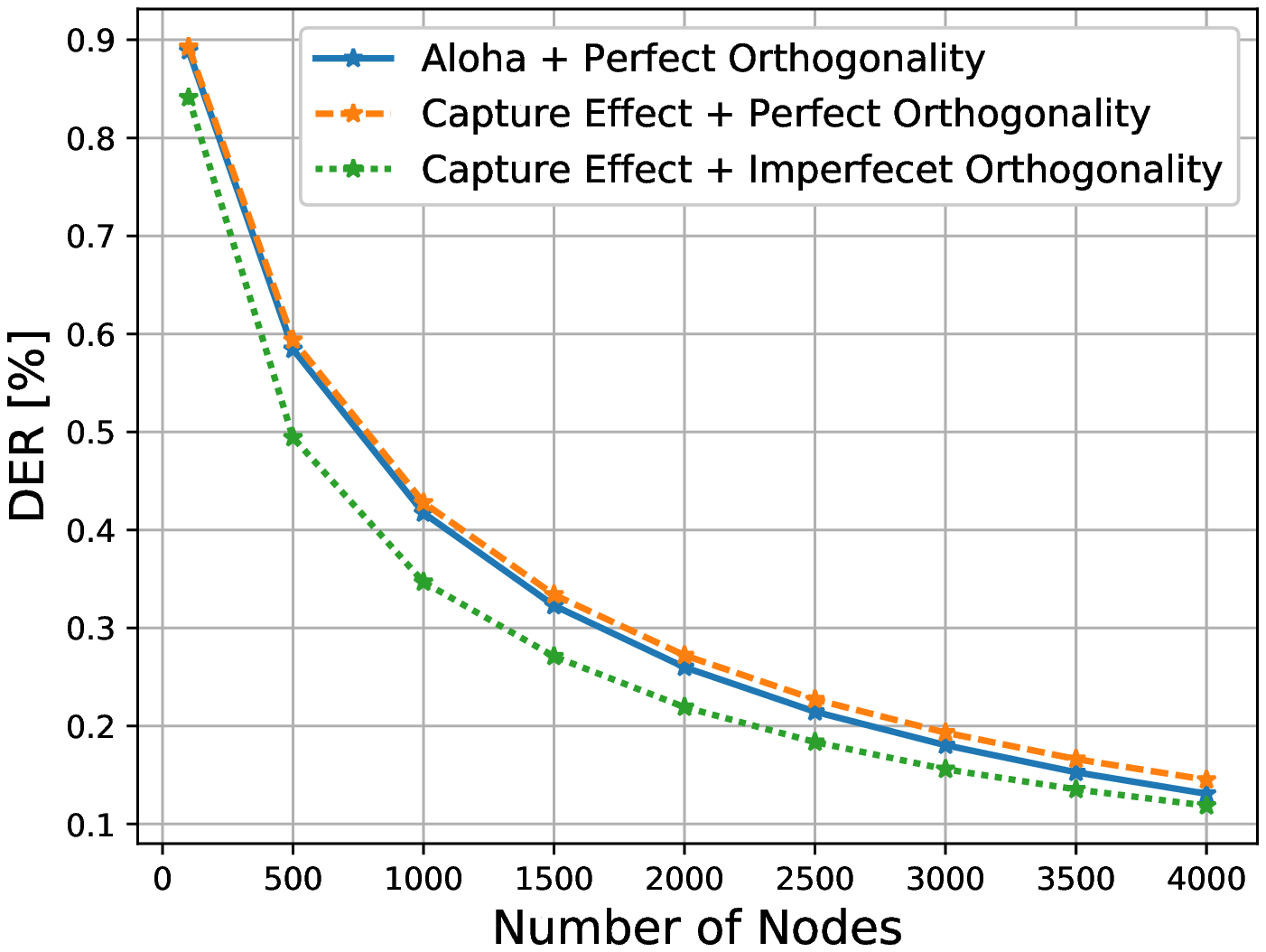}
  }
  \vspace{-1em}
  \caption{Study of Aloha, Capture and Imperfect Orthogonality effects}
  \vspace{-0.5cm}
  \label{fig:study}
\end{figure}


As LoRaWAN is based on Aloha, it is supposed to be a fair MAC protocol.
However, the above effects introduce unfairness, favouring transmissions from nodes closer to the gateway and by those that use lower spreading factors. 
The impact of these issues was validated experimentally, with all nodes allocated the same transmission power level (TP), and the SFs allocated uniformly, in the order of distance from the gateway. Figure~\ref{fig:fairness_study} shows Jain's fairness index,
$\zeta = \frac{(\sum_{i=1}^n DER_i)^2}{n\sum_{i=1}^n DER_i^2}$,
where $DER$ denotes the Data Extraction Rate. When all the issues are considered the fairness decreases drastically with increasing network size. Even without capture effect and with perfectly orthogonal spreading codes, the fairness of Aloha decreases as well, due to the different collision probabilities of the SFs. 

Figure~\ref{fig:der_study} shows the DER and as expected the capture effect favours the nodes closest to the gateway, which is reflected in the overall DER of the system. However, considering the imperfect orthogonality effect, the overall DER is lower than Aloha. Here, we propose an approach to remedy this. Our proposal, FADR, a Fair Adaptive Data Rate algorithm for LoRaWAN, computes the optimal combination of SFs and TPs settings to achieve a fair DER among all nodes.
%

\begin{figure*}
\vspace{-2.1em}
  \centering
  \subfloat[Fairness Index]{\label{fig:res_fair}\includegraphics[width=0.24\textwidth]{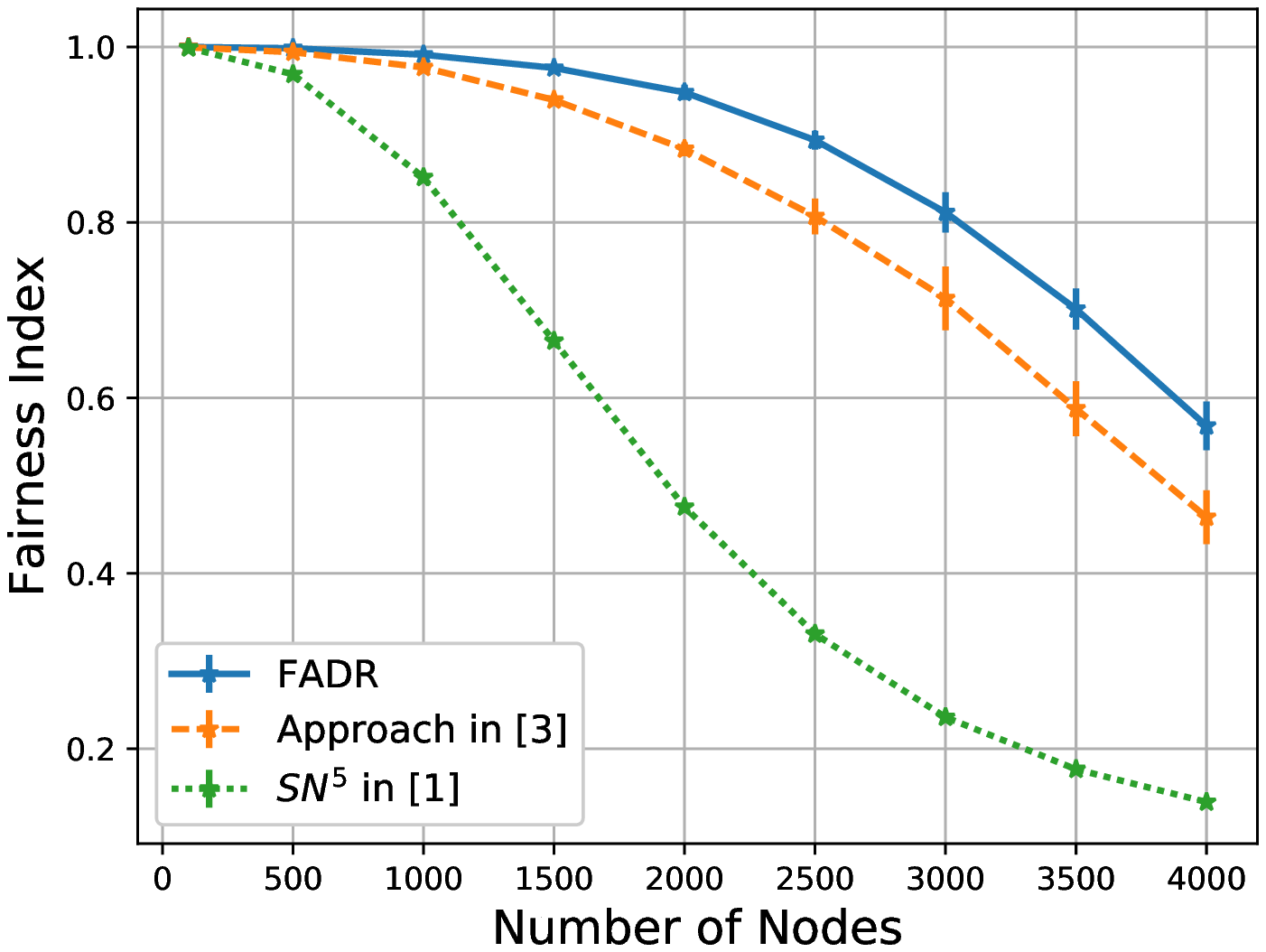}}
  \subfloat[Overall DER]{\label{fig:res_der}\includegraphics[width=0.24\textwidth]{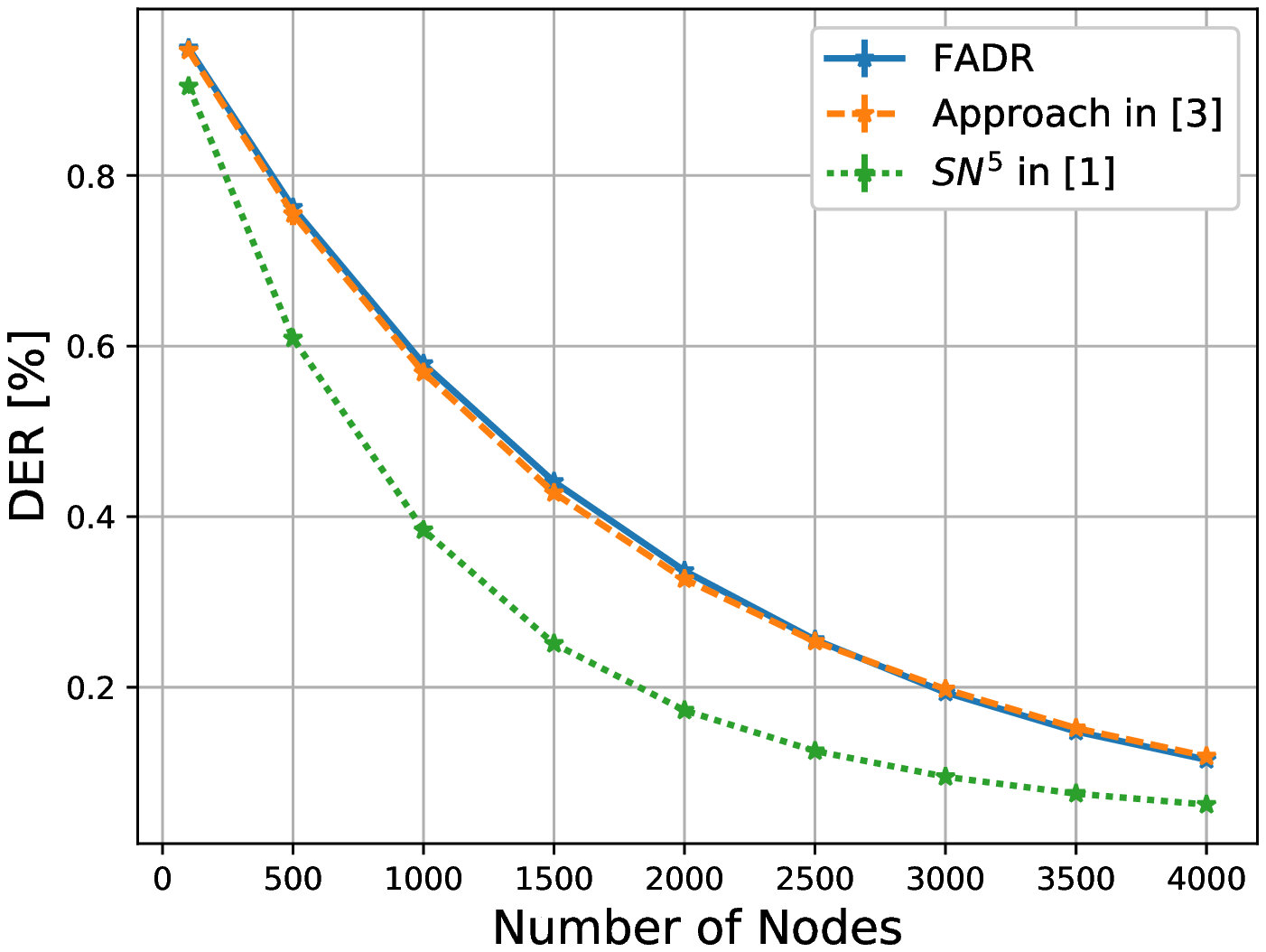}}
  \subfloat[Overall Energy]{\label{fig:energy}\includegraphics[width=0.24\textwidth]{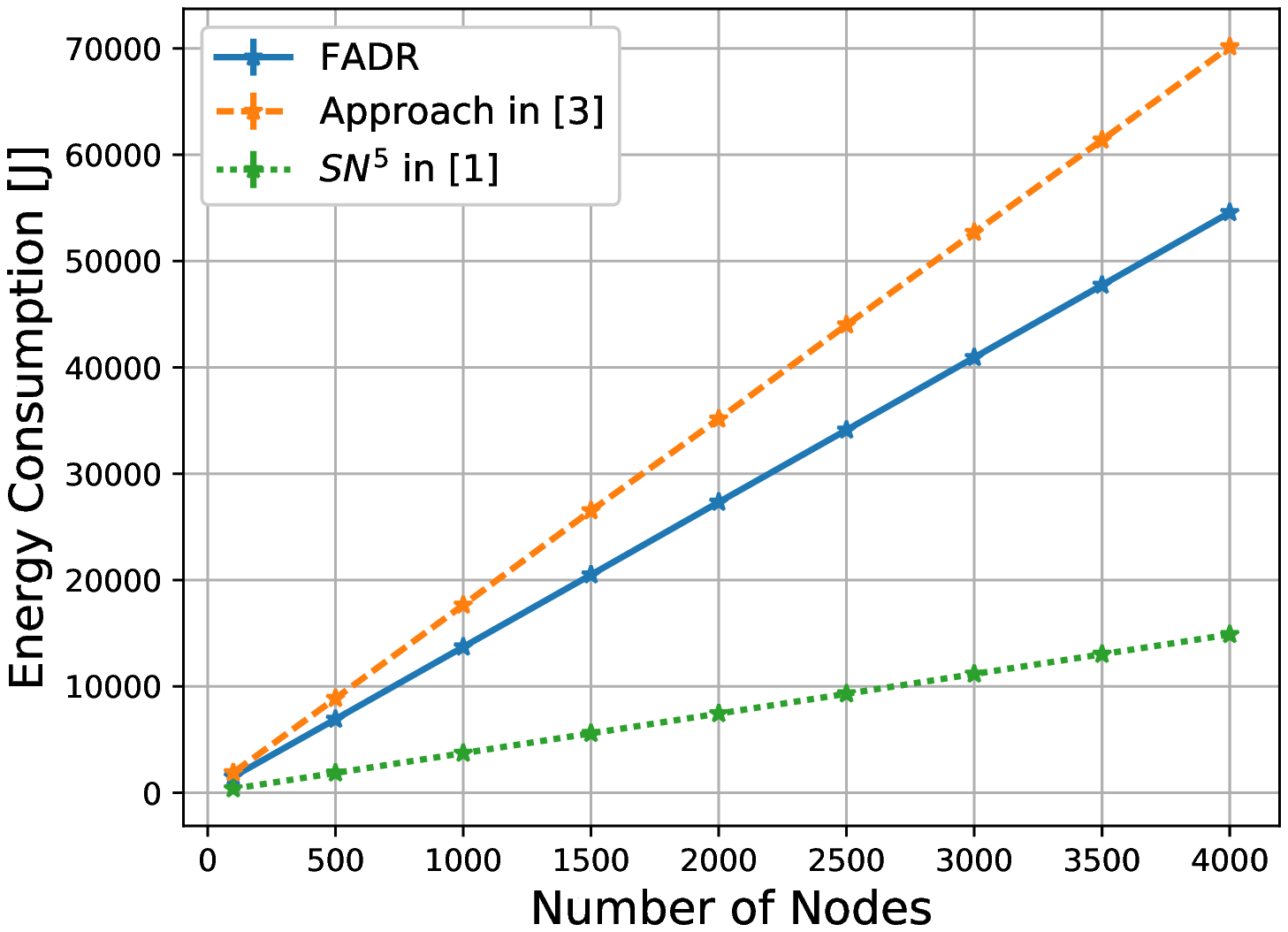}}
  \subfloat[DER vs distance]{\label{fig:res_der_dist}\includegraphics[width=0.24\textwidth]{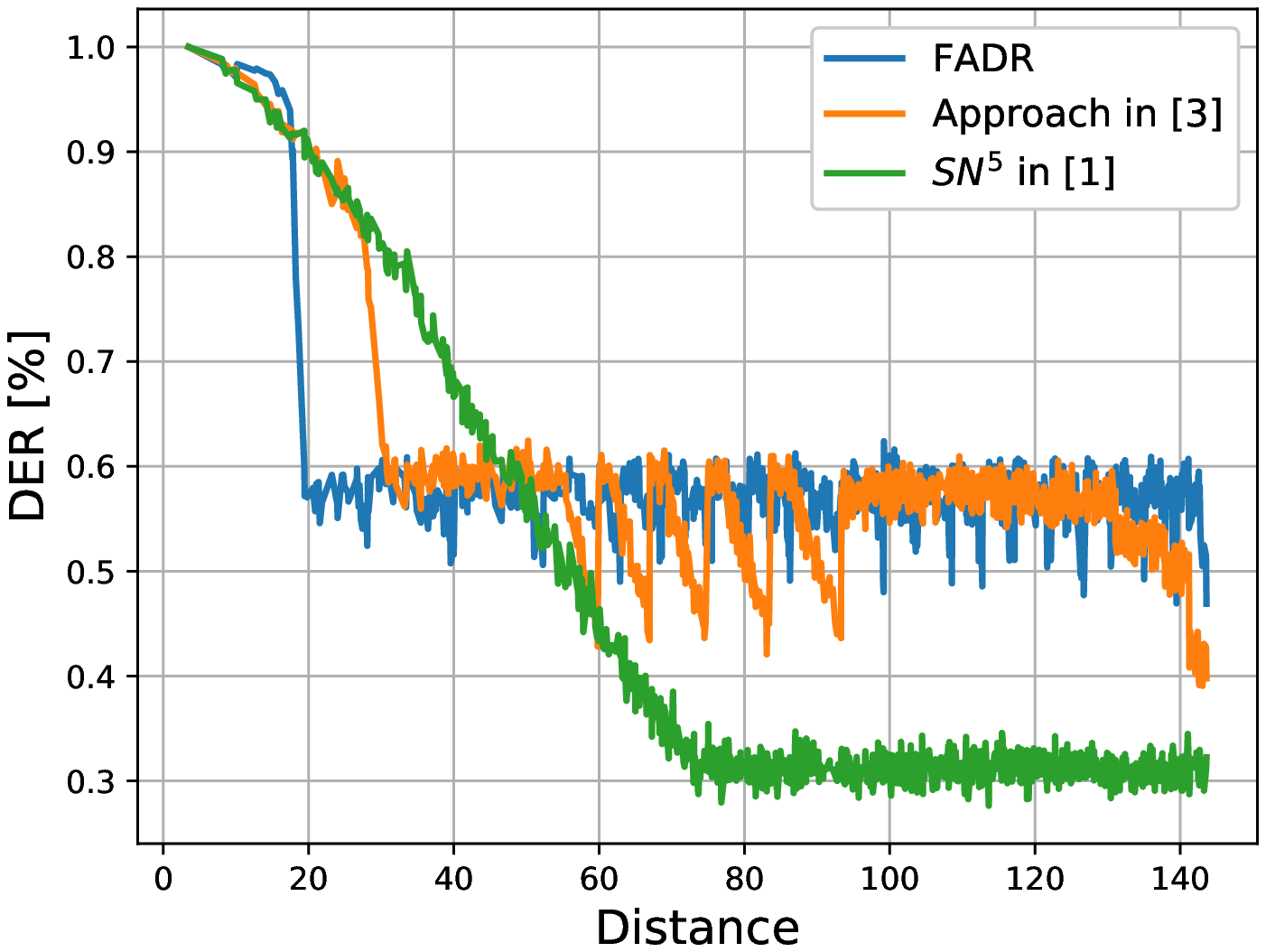}}
  \vspace{-1em}
  \caption{Evaluation results}
  \vspace{-0.5cm}
  \label{fig:results}
\end{figure*}

\begin{algorithm}
  \caption{FADR - Power Allocation Algorithm}
  \label{alg:fadr}
  \begin{algorithmic}[1]
    \small
    \STATE \textbf{Input} List of nodes \textbf{N}, corresponding \textbf{RSSI}, power levels \textbf{PowLevels}, matrix of \textbf{CIR}
    \STATE \textbf{Output} $\forall n \in \textbf{N}, \textbf{P}[n] \in \textbf{PowLevels}$
    \STATE Sort \textbf{N} by \textbf{RSSI}
    \STATE $RSSI_m = min(\textbf{RSSI})$, $RSSI_M = max(\textbf{RSSI})$, $CIR_m=min(\textbf{CIR})$
    \STATE $Pow_m \leftarrow \textbf{PowLevels}.pop(0)$
    \STATE $Pow_M \leftarrow \argmin_{i\in\textbf{PowLevels}} |RSSI_M + Pow_m - RSSI_m - i| \le CIR_m$
    \STATE $\textbf{PowLevels} = [0:\textbf{PowLevels}.index(Pow_M)]$
    \STATE $RSSI_m \leftarrow min(RSSI_m+Pow_M,RSSI_M+Pow_m)$
    \STATE $RSSI_M \leftarrow max(RSSI_m+Pow_M,RSSI_M+Pow_m)$
    \STATE $i_0 \leftarrow \argmax_{i<|N|} |\textbf{RSSI}[i]+Pow_m|\le|RSSI_m|$
    \STATE $\textbf{P}[i] = Pow_m \forall i < i_0$
    \STATE $i_n \leftarrow \argmax_{|N|>i>i_0} |\textbf{RSSI}[i] + Pow_M - RSSI_m| > CIR_m$
    \STATE $\textbf{P}[i] = Pow_M \forall i > i_n$
    \STATE $idx \leftarrow i_0 + 1$
    \FORALL{$p \in \textbf{PowLevels}$}
    \IF{$|\textbf{RSSI}[idx] + p - RSSI_m|\le CIR_m$ \AND $|\textbf{RSSI}[idx] + p - RSSI_M - Pow_M| \le CIR_m$}
    \STATE $i_k \leftarrow \argmax_{idx < j < i_n} |\textbf{RSSI}[j] + p - \textbf{RSSI}[i_n] - Pow_M| \le CIR_m$
    \STATE $\textbf{P}[j] = p \forall j \in [idx,i_k]$, $idx = i_k$
    \ENDIF
    \ENDFOR
  \end{algorithmic}
\end{algorithm}

\section{The FADR Algorithm}
\label{sec:fadr}
The FADR algorithm manages the allocation of SFs and TPs to the nodes. For SF allocation we use the optimal SF distribution for fair collision probability, determined in~\cite{reynders2017power}. However, while the authors in ~\cite{reynders2017power} apply this to the entire network, based on distance from the gateway, we propose to assign it over regions. To this extent,
%
the nodes are first ordered based on RSSI and divided into groups of 50 to overcome the rounding problem and for better representation of all SFs in which SF 12 is assigned to one node. The optimal SF distribution is applied to each group of 50 nodes.

The allocation of transmission power levels is shown in algorithm~\ref{alg:fadr}. The algorithm allocates the lowest TPs that can reduce the difference in RSSI below the CIR threshold to mitigate the capture effect and the imperfect orthogonality of spreading codes.
FADR assumes that all nodes are initiated with the same transmission power. FADR requires the RSSI corresponding to each node, so it is run only after a certain number of packets have been collected by the network.


\section{Evaluation and Results}
\label{sec:eval}

FADR was implemented in a version of LoRaSim~\cite{bor2016lora} modified to include the effect of imperfect orthogonality of SFs. FADR was compared with the state of the art in~\cite{reynders2017power}, and with the minimum transmission time algorithm $SN^5$ in~\cite{bor2016lora}, from the point of view of DER fairness and overall energy consumption. To replicate the evaluation in~\cite{reynders2017power} the minimum sensitiviy of all SFs in LoRaSim was lowered to -140 dBm, so that all nodes can reach the gateway with all combinations of SF and TP. The CIR and capture effect thresholds were set to 6dB for all SFs, based on the work of~\cite{croce2017impact} and \cite{bor2016lora}, respectively. The number of nodes was varied from 100 to 4000 nodes, packet length set to 80 bytes and average packet transmission interval to 1 minute. Each experiment was run for a simulation time of one day and repeated with 10 different random seeds.

The $SN^5$ algorithm assigns to nodes the lowest SF and the lowest available TP.
The algorithm in \cite{reynders2017power} assigns SFs based on a distribution that achieves a fair probability of collisions (FADR uses the same). Allocating over the whole network will result in almost 45\% of nodes at SF 7 and the same TP, which increases the probability of the capture effect. The TP control algorithm attempts to minimise RSSI difference, however the reference is the node with the highest path loss and SF 8, assuming that this node is having the highest error rate. 
This assumption is valid only in uniformly distributed networks, because the location and path loss of this node depends on the node placement. In a large network this node could be close to the sink and with low path loss; in a small network it might be closer to the edge, with high path loss.

The results in figure~\ref{fig:results} show that FADR surpasses the state of the art~\cite{reynders2017power} in fairness and energy consumption, without sacrificing the DER. At 4000 nodes, FADR achieves 300\% higher fairness than \cite{bor2016lora} and 22\% higher than \cite{reynders2017power}. FADR assigns lower TPs than \cite{reynders2017power}, achieving 22\% reduction in energy consumption. However, both consume more energy than \cite{bor2016lora} in which all nodes transmit with 
the lowest TP. FADR's advantage over \cite{reynders2017power} is shown in figure~\ref{fig:res_der_dist} (1000 nodes). 
FADR achieves roughly the same DER for a larger proportion of the network than \cite{reynders2017power}, which experiences high variation for nodes with SF 7 and high path loss, thus, FADR is fairer.

\section{Conclusions}
We propose FADR to achieve a fair data extraction rate in LoRa/LoRaWAN by exploiting optimal combinations of spreading factors and transmission power levels and at the same time maintain node lifetime by not using excessively high transmission power levels. Simulations show that FADR, in a highly congested cell, achieves 300\% higher fairness than the minimum airtime allocation approach and 22\% higher fairness than Brecht’s approach, which is one of the state-of-art approaches, targeting the same problem with almost 22\% lower network energy consumption. For future work, we will implement FADR in a real LoRa deployment.

\textit{\textbf{Acknowledgments}} - this research has received support from Science Foundation Ireland (SFI) under Grant Number 13/RC/2077.
%
%

%

\bibliographystyle{abbrv}
\bibliography{main}  
\end{document}